\documentclass[prb,preprint]{revtex4} 
\usepackage{amsmath}  % needed for \tfrac, \bmatrix, etc.
\usepackage{amsfonts} % needed for bold Greek, Fraktur, and blackboard bold
\usepackage{graphicx} % needed for figures

\begin{document}

\title{Simple pendulum dynamics: revisiting the Fourier-based approach to the solution}

\author{Riccardo Borghi}
\email{borghi@uniroma3.it}
\affiliation{Dipartimento di Elettronica Applicata, Universit\`a degli Studi ``Roma tre''\\
Via della vasca navale 84, I-00144 Rome, Italy}

\begin{abstract}
The Fourier-based analysis customarily employed to analyze 
the dynamics of a simple pendulum is here revisited to propose
an elementary iterative scheme aimed at generating a sequence of 
analytical approximants of the exact law of motion. Each approximant is 
expressed by a Fourier sum whose coefficients are given by suitable 
linear combinations of Bessel functions, which are expected to be more accessible, especially at an undergraduate level, with respect to Jacobian elliptic functions. 
The first three approximants are explicitely obtained and 
compared with the exact solution for typical initial angular 
positions of the pendulum. In particular, it is shown that,
at the lowest approximation level, the law of motion of the pendulum 
turns out to be adequately described, up to oscillation amplitudes of 
$\pi/2$, by a sinusoidal temporal behaviour with a frequency proportional 
to the square root of the so-called ``besinc'' function, well known in physical optics. 
\end{abstract}

\maketitle

%%%%%%%%%%%%%%%%%%%%%%%%%%%%%%%%%%%%%%%%%%%%%%%%%%%%%%%%%%%%%%%

\section{Introduction}
\label{Sec:Introduction}

% RECENT WORKS ON SIMPLE PENDULUM

The simple pendulum is the first example, met at an 
undergraduate level, of a nonlinear oscillator whose temporal law
of motion does not have a solution expressible in terms of elementary 
functions~\cite{Sommerfeld/1970}, being it customarily given via the use of 
the so-called Jacobian elliptic functions~\cite{DLMF}, not yet available 
in the ``box of tools''~\cite{Feynamn/Leighton/1985} of first-year physics students. As a consequence, elementary analyses of the pendulum dynamics are generally limited to ``small'' oscillations, for which the differential equation can be linearized to predict an approximately harmonic law of motion along the pendulum circular trajectory. Such isochronous solution can be intrepreted, through the joint use of a perturbative approach originally proposed by Lord Rayleigh~\cite{Rayleigh/1899,Minorsky/1947,Lamb/1961} and of a Fourier series expansion, as the lowest-order term of a sequence of analytical approximants of the exact solution that can be generated iteratively. This has been done by Fulchner and Davis~\cite{Fulchner/Davis/1975}, which provided polynomial approximants, with respect to the oscillation amplitude, of the Fourier coefficients of the first three elements of the sequence. The results obtained in~\cite{Fulchner/Davis/1975} have been used in some of the several papers published for nearly thirty years to give new and original insights, both theoretical and experimental, about the pendulum dynamics~\cite{Simon/Riesz/1979,Zilio/1981,Ganley/1985,Cadwell/1991,Molina/1997,Kidd/2002,Millet/2003,Parwani/2004,%
Hite/2005,Belendez/2006,Lima/Arun/2006,Belendez/2007,Amore/2007,Gil/Legarreta/DiGregorio/2008,%
Carvalhaes/2008,Lima/2008,Belendez/2009,Quing-Xin/Pei/2009,Belendez/2010,Turkyilmazoglu/2010,%
Qureshi/2010,Ochs/2011,Belendez/2011,Douvropoulos/2012,Sheng/2012}.

The aim of the present paper is to provide further insights to the Fourier-based treatment of the pendulum dynamics. Following the Rayleigh's path, we 
also propose an iterative approach to solve the nonlinear equation 
in terms of truncated Fourier series arranged in a sequence of steps 
of growing complexity. However, differently from~\cite{Fulchner/Davis/1975}, the Fourier expanding coefficients are not expressed via polynomial approximants but rather obtained, through suitable orthogonal expansions, in terms of suitable linear combinations of Bessel functions~\cite{DLMF}. 
A quantitative comparison with the exact solution is then carried out for the first three iterations and shows that, at the lowest approximation level the temporal law of motion of the pendulum turns out to be adequately described for oscillation amplitudes up to $\pi/2$ (with accuracies better than 0.1 \% as far as the pendulum period is concerned) by the so-called ``besinc'' function, 
well known in optics where it describes mathematically the light distribution produced via diffraction by circular holes~\cite{Born/Wolf/1999}.

\section{Theoretical analysis}
\label{Sec:TheoreticalAnalysis}
\subsection{The Fourier analysis}
\label{SubSec:FourierAnalysis}

% THE H LAMB APPROACH
We denote by $\vartheta=\vartheta(t)$ the angular position of a simple pendulum (with respect to the vertical direction)
as a function of the time $t$. Newton's law gives for $\vartheta$ the well known  nonlinear differential equation  
\begin{equation}
\label{Eq:Lamb.1}
\displaystyle
\ddot{\vartheta}\,+\,\omega^{2}_{0}\,\sin\vartheta\,=\,0\,,
\end{equation}
with $\omega_{0}$ denoting the natural frequency of the pendulum  oscillations %in the limit of small amplitudes  
and the dots derivation with respect to $t$. We consider symmetric (with respect to the vertical direction 
$\vartheta=0$) oscillations  having  amplitude  $\vartheta_0$,  so that Eq.~(\ref{Eq:Lamb.1}) must be solved together with the initial conditions
\begin{equation}
\label{Eq:Lamb.2}
\begin{array}{l}
\displaystyle
\vartheta(0)\,=\,\vartheta_0\,,\\
\\
\displaystyle
\dot{\vartheta}(0)\,=\,0\,.
\end{array}
\end{equation}
Harmonic solutions of this equation are achieved in the limit of small amplitudes (i.e., for $|\vartheta| \ll 1$) which, via the 
 approximation $\sin\vartheta \simeq \vartheta$, transforms Eq.~(\ref{Eq:Lamb.1}) into 
\begin{equation}
\label{Eq:Lamb.4}
\displaystyle
\ddot{\vartheta}\,+\,\omega^{2}_{0}\,\vartheta\,=\,0\,,
\end{equation}
whose integral, satisfying initial conditions~(\ref{Eq:Lamb.2}), turns out to be
\begin{equation}
\label{Eq:Lamb.5}
\displaystyle
\vartheta(t)\,=\,\vartheta_0\,\cos \omega_{0} t\,,
\end{equation}
representing isochronous oscillations with period $T_0=2\pi/\omega_{0}$. For amplitude values beyond  the  above approximation, the pendulum oscillations are no longer isochronous and the corresponding period, say $T$, display the following (exact) dependance on $\vartheta_0$~\cite{Lamb/1961}:
\begin{equation}
\label{Eq:Lamb.7}
\displaystyle
\frac{T}{T_0}\,=\,\frac 2\pi\,K\left(\sin\frac{\theta_0}2\right)\,,
\end{equation}
where $K(\cdot)$ denotes the complete elliptic integral of the first kind~\cite{DLMF}. Moreover, the exact law of motion $\vartheta=\vartheta(t)$ is given by
\begin{equation}
\label{Eq:Lamb.16.1}
\displaystyle
\vartheta(t)\,=\,2\,\arcsin\left\{ \sin\frac{\vartheta_0}2\,\mathrm{sn}\left[\frac{\omega_0}{\omega}\left(\frac \pi 2\,-\,\omega t\right)\,\left|\,\, \sin\frac{\vartheta_0}2\right.\right]\right\} \,,
\end{equation}
where the symbol $\mathrm{sn}(\cdot | \cdot)$ denotes the Jacobian elliptic function~\cite[Ch.~22]{DLMF} and $\omega=2\pi/T$. 
Equation~(\ref{Eq:Lamb.16.1}) is periodic with respect to $t$ with period
$T$, so that it can be written as a Fourier series. Due to the parity of the function $\sin\vartheta$ in the r.h.s. of Eq.~(\ref{Eq:Lamb.1}), only odd
multiples of the fundamental frequency $\omega$ will appear in the Fourier
expansion of $\vartheta$ which, on taking Eq.~(\ref{Eq:Lamb.2}) into account,
reads
\begin{equation}
\label{Eq:Lamb.12}
\displaystyle
\vartheta(t)\,=\,
\sum_{k=0}^\infty\,
c_{2k+1}\,\cos\,[(2k+1)\omega t]\,,
\end{equation}
where the expanding coefficients $c_k$ ($k=1,3,5,...$) are functions of
the sole initial amplitude oscillation $\vartheta_0$.  
Differently from the exact expression
given in Eq.~(\ref{Eq:Lamb.16.1}), the Fourier representation in 
Eq.~(\ref{Eq:Lamb.12}) presents a more transparent mathematical 
structure from which the role played by the various expanding coefficients in contributing to give the temporal law of motion its temporal ``shape'' can be inferred. In particular, for ``small'' initial positions all coefficients
but the first will turn out to be negligible, being at the same time 
$c_1\simeq\vartheta_0$. 
%On the contrary, for $\vartheta_0\to\pi$ we should
%expect $\vartheta(t)\to\pi$, which is actually a solution of
%Eqs.~(\ref{Eq:Lamb.1}) with an infinite period of oscillation.
All Fourier coefficients $c_k$ could be, in principle, obtained starting
from the analytical form of the solution given in Eq.~(\ref{Eq:Lamb.16.1})
through the following expression:
\begin{equation}
\label{Eq:Lamb.12.1}
\begin{array}{l}
\displaystyle
c_{2k+1}\,=\,\frac 2{T}\,
\int_0^{T}\,\vartheta(t)\,\cos\,[(2k+1)\omega t]\,\mathrm{d}t\,=\,\\
\\
\displaystyle
\,=\,
\frac 2\pi\,
\int_0^{2\pi}
\arcsin\left\{ \sin\frac{\vartheta_0}2\,\mathrm{sn}\left[\frac{\omega_0}{\omega}\left(\frac \pi 2\,-\,\xi\right)\,\left|\,\, \sin\frac{\vartheta_0}2\right.\right]\right\} \,
\,\cos\,[(2k+1)\xi]\,\mathrm{d}\xi\,,
\end{array}
\end{equation}
($k=0,1,2,\ldots$) which, although cannot be given a closed form, are easily attainable via standard numerical packages.{In the present paper all numerical calculations have been performed via the commercial software 
{\tt Mathematica 8.0}.}
The problem of giving analytical estimates of the above expanding coefficients
has been tackled by Fulchner and Davis~\cite{Fulchner/Davis/1975}, which provided Taylor-like polynomial approximants, with respect to $\vartheta_0$, of $c_1$, $c_3$, and $c_5$. 

\subsection{The lowest-order solution}
\label{SubSec:LowestOrderSolution}

In the present paper we show further analytical approximants of the coefficients~(\ref{Eq:Lamb.12.1}) 
that can be obtained by solving Eq.~(\ref{Eq:Lamb.1}) at a different levels of approximations, each of 
them obtainable from the previous one in an iterative manner. To this end we start on substituting from 
Eq.~(\ref{Eq:Lamb.12}) into  Eq.~(\ref{Eq:Lamb.1}) so that
\begin{equation}
\label{Eq:Lamb.14}
\begin{array}{l}
\displaystyle
\omega^2\,(c_1\,\cos\omega t\,+\,9\,c_3\,\cos 3\omega t\,+\,\ldots)\,=\,\\
\\
\displaystyle
\,=\,
\omega^2_0\,\sin(c_1\,\cos\omega t\,+\,c_3\,\cos 3\omega t\,+\,\ldots)\,,
\end{array}
\end{equation}
and require the expanding coefficients $\{c_{2k+1}\}$  and the frequency $\omega$ 
to be found simply on imposing that the periodic functions in both sides of 
Eq.~(\ref{Eq:Lamb.14}) have the same Fourier representation, together with the supplementary constraint
\begin{equation}
\label{Eq:Lamb.14.1}
\begin{array}{l}
\displaystyle
c_1\,+\,c_3\,+\,...\,=\,\vartheta_0\,,
\end{array}
\end{equation}
which directly arises from the initial conditions in Eq.~(\ref{Eq:Lamb.2}). 
We are going to solve Eq.~(\ref{Eq:Lamb.14}) by successive approximation,
starting from the lowest-order solution which corresponds to the choice
$c_1\,\simeq \vartheta_0$ and $c_3\,=\,c_5\,=\,\ldots\,=\,0$, 
so that we have
\begin{equation}
\label{Eq:Lamb.15}
\begin{array}{l}
\displaystyle
\omega^2\,\vartheta_0\,\cos\omega t\,\simeq
\omega^2_0\,\sin(\vartheta_0\,\cos\omega t)\,,
\end{array}
\end{equation}
where the function in the r.h.s. can be expanded as a Fourier series by 
using formula~(10.12.3) of Ref.~\cite{DLMF} to obtain
\begin{equation}
\label{Eq:Lamb.15.0}
\begin{array}{l}
\displaystyle
\sin(\vartheta_0\,\cos\omega t)\,=\,
2\,\sum_{k=0}^\infty\,
(-1)^k\,J_{2k+1}(\vartheta_0)\,\cos\,[(2k+1)\omega t]\,=\,\\
\\
\displaystyle
\,=\,2\,J_1(\vartheta_0)\,\cos\omega t\,-\,
2\,J_3(\vartheta_0)\,\cos 3\omega t\,+\,\ldots\,,
\end{array}
\end{equation}
with $J_n(\cdot)$ denoting the $n$th-order Bessel function of the first
kind~\cite{DLMF}. 
In particular, to find the lowest-order estimate of the frequency $\omega$
it is sufficient to keep in the Fourier expansion~(\ref{Eq:Lamb.15.0}) only the first term which, once substituted into Eq.~(\ref{Eq:Lamb.15}), gives at once
\begin{equation}
\label{Eq:Lamb.15.1}
\begin{array}{l}
\displaystyle
\frac{\omega^2}{\omega^2_0}\,\simeq\,
\frac{2\,J_1(\vartheta_0)}{\vartheta_0}\,.
\end{array}
\end{equation}
{Alternatively we can multiply both sides of Eq.~(\ref{Eq:Lamb.15}) by 
$\cos\omega t$, integrate with respect to the dimensionless variable
$\xi=\omega t$ within the interval $[0,2\pi]$, and recall the integrals
\[
\begin{array}{l}
\displaystyle
\int_0^{2\pi}\,\cos^2\xi\,\mathrm{d}\xi\,=\,\pi\,,\\
\\
\displaystyle
\int_0^{2\pi}\,\cos \xi\,\sin(\vartheta_0\,\cos\xi)\mathrm{d}\xi\,=\,
2\pi\,J_0(\vartheta_0)\,,
\end{array}
\]
}
Before proceeding it worth noting that the function in the r.h.s. of 
Eq.~(\ref{Eq:Lamb.15.1}) is exactly the so-called ``besinc'' function,
defined as
\begin{equation}
\label{Eq:BesincFunction}
\mathrm{besinc}(\alpha)\,=\,\frac{2\,J_1(\alpha)}{\alpha}\,,
\end{equation}
well known in diffraction theory, where describes mathematically the characteristic figure produced by light via diffraction through circular holes, the so-called ``Airy disk''~\cite{Born/Wolf/1999}.
\begin{figure}[!ht]
 \begin{minipage}[b]{6.cm}
 \centerline{
   \includegraphics[width=6cm]{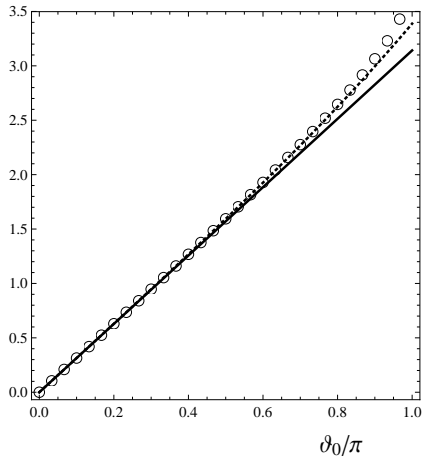}%{Fig.C1.Lowest.eps}
   }
 \end{minipage}
\caption{Behaviour, as a function of $\vartheta_0$, of the Fourier expanding coefficient $c_1$
numerically evaluated from Eq.~(\ref{Eq:Lamb.12.1})  (open circles), together with its  lowest-order
linear approximation (solid line). For completeness the polynomial approximants of $c_1$ provided in~\cite{Fulchner/Davis/1975}
is also shown (dotted curve)}
\label{Fig:Lamb.0}
\end{figure}
Figure~\ref{Fig:Lamb.0} shows the comparison between the exact values (open circles) of the expanding coefficient $c_1$, obtained by numerically evaluating the r.h.s of Eq.~(\ref{Eq:Lamb.12.1}) for $k=0$, and the  lowest-order linear approximation (solid line), for $\vartheta_0\in[0,\pi]$. For completeness the polynomial approximants of $c_1$ provided in~\cite{Fulchner/Davis/1975} is also shown (dotted curve). To give an idea to the reader about the limits of the 
 lowest-order approximation  of $\vartheta(t)$, i.e.,  
\begin{equation}
\label{Eq:Lamb.17}
\displaystyle
\vartheta(t)\,\simeq \,\vartheta_0\,\cos\omega t\,,
\end{equation}
with $\omega$ being given by Eq.~(\ref{Eq:Lamb.15.1}),
Fig.~\ref{Fig:Lamb.2} shows, for different values of $\vartheta_0$, a visual comparison between the temporal behaviours 
of the exact (open circles) law of motion in Eq.~(\ref{Eq:Lamb.16.1}), the  lowest-order estimate in Eq.~(\ref{Eq:Lamb.17})~(solid curve), and the harmonic solution~(dotted curve) given in Eq.~(\ref{Eq:Lamb.5}).
\begin{figure}[!ht]
 \begin{minipage}[b]{6cm}
   \centering
   \includegraphics[width=6cm]{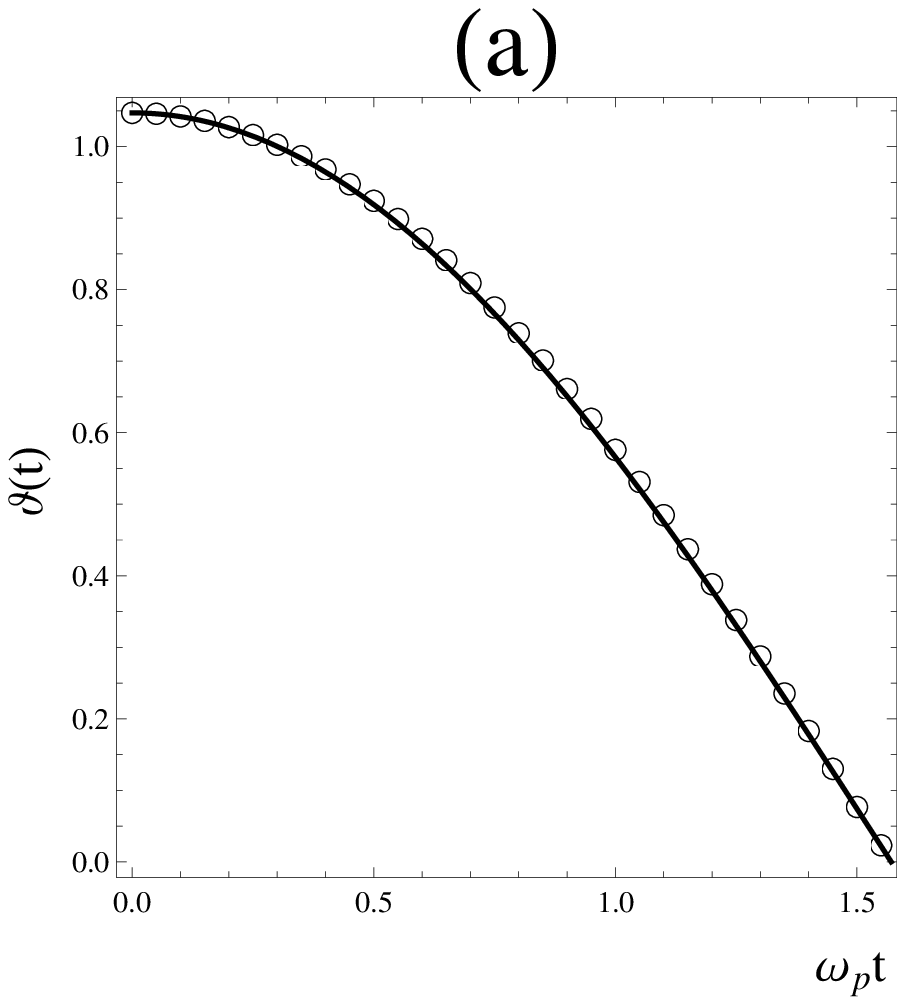} %{Fig.Comparison.LowestA.eps}
   \includegraphics[width=6cm]{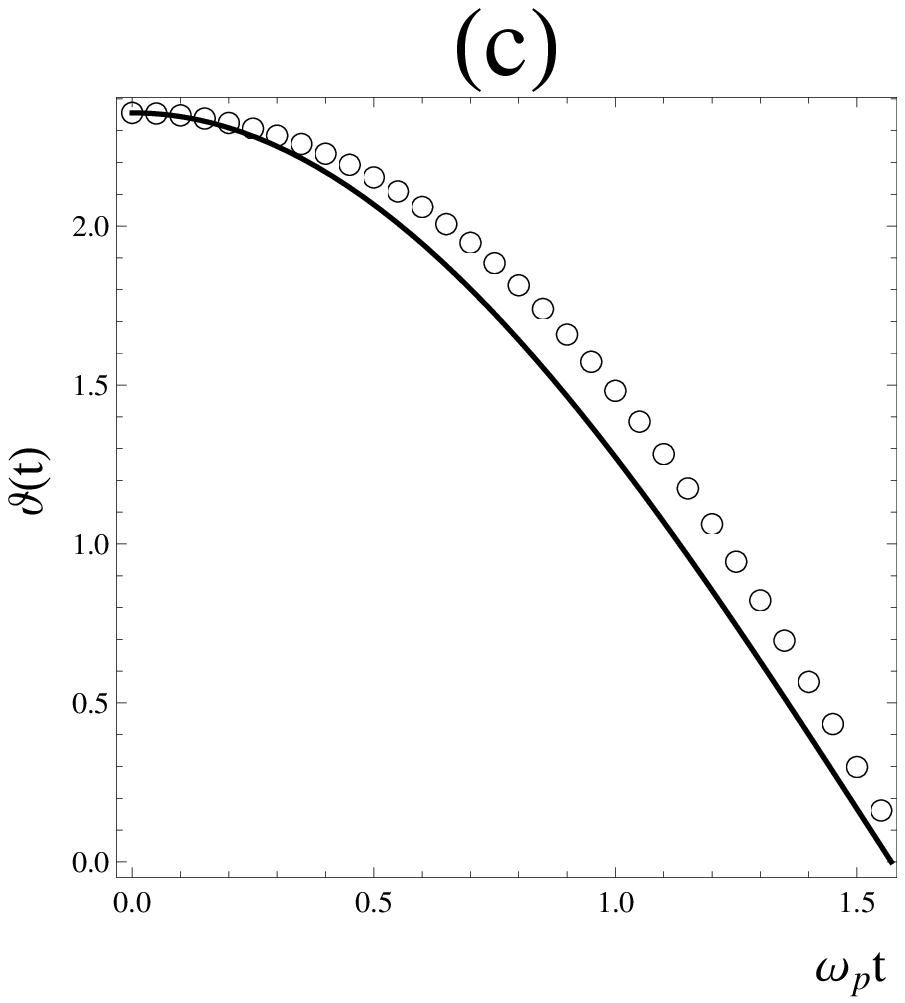} %{Fig.Comparison.LowestC.eps}
 \end{minipage}
 \ \hspace{2mm} \hspace{3mm} \
 \begin{minipage}[b]{6cm}
  \centering
   \includegraphics[width=6cm]{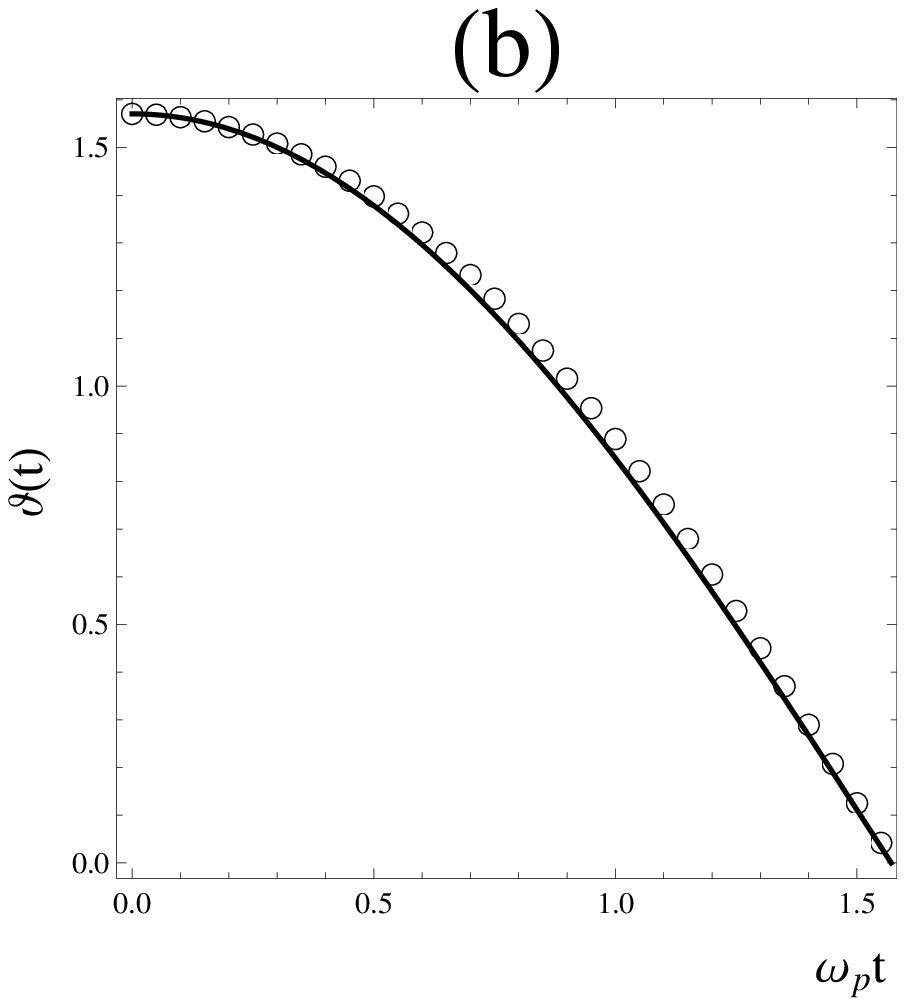} %{Fig.Comparison.LowestB.eps}
   \includegraphics[width=6cm]{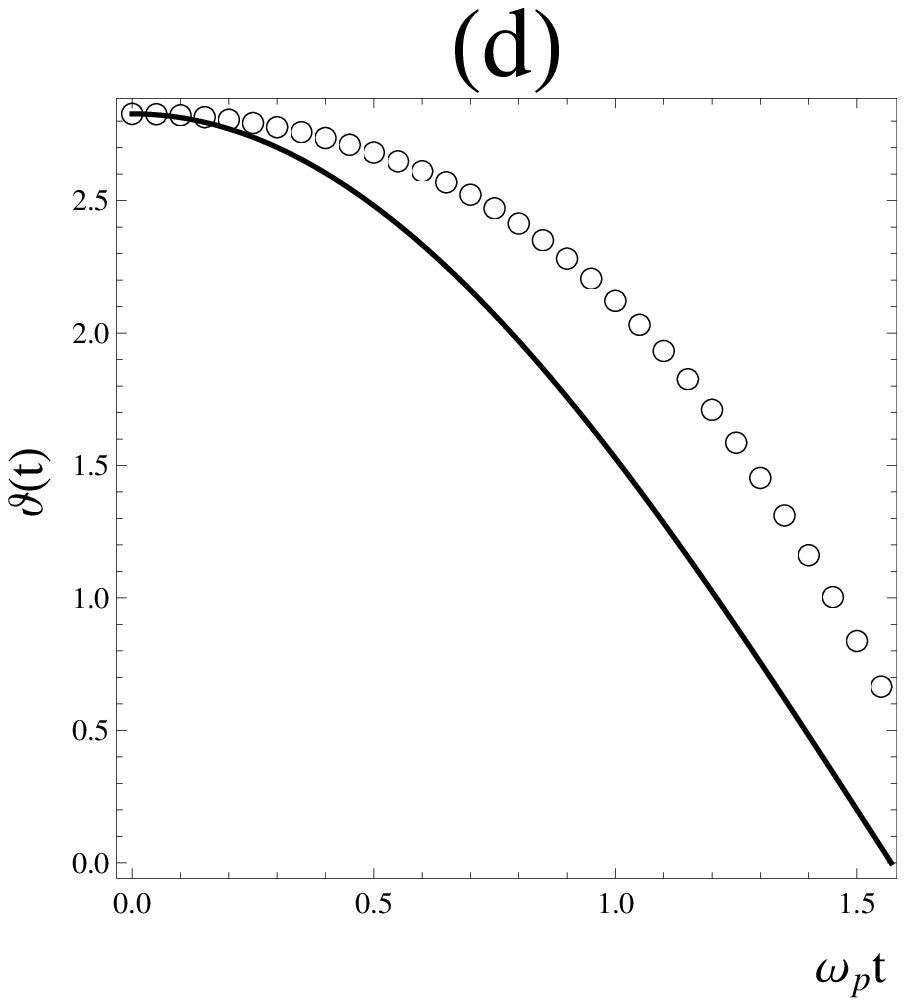} %{Fig.Comparison.LowestD.eps}
 \end{minipage}
\caption{Visual comparison between the temporal behaviours of the exact 
(open circles) law of motion, the lowest-order estimate in Eq.~(\ref{Eq:Lamb.17})~(solid curve), and the harmonic solution 
in Eq.~(\ref{Eq:Lamb.5})~(dotted curve),
for $\vartheta_0=\pi/3$ (a), $\pi/2$ (b), $3\pi/4$ (c), and $9\pi/10$ (d). }
\label{Fig:Lamb.2}
\end{figure}
Figure~\ref{Fig:Lamb.2}(a),  corresponding to $\vartheta_0=\pi/3$, displays an optimum agreement with the exact solution, as well as  
Fig.~\ref{Fig:Lamb.2}(b)  ($\vartheta_0=\pi/2$), for which  the agreement, at least at a visual level, could be considered acceptable. 
Clearly the same does not happen for  figures~(c) and~(d), corresponding  to $\vartheta_0=3\pi/4$ and $9\pi/10$, respectively, 
for which a deeper investigation is required.

\subsection{Higher-order solutions}
\label{SubSec:FirstOrderSolution}

The derivation of the lowest-order solution can be further refined in the following way: the first-order estimate of 
$\vartheta(t)$ will obtained by truncating the Fourier series in Eq.~(\ref{Eq:Lamb.12}) up to $k=1$, i.e., by
including now the first two harmonics. To find the corresponding coefficients, say $\{c'_1,c'_2\}$, and also the new
estimate of the frequency we can substitute  the lowest-order solution given in Eq.~(\ref{Eq:Lamb.17})
in the r.h.s. of Eq.~(\ref{Eq:Lamb.14}) and to expand it, thanks to the Fourier series expansion in Eq.~(\ref{Eq:Lamb.15.0}), up to the term
containing $\cos 3\omega t$ to obtain
\begin{equation}
\label{Eq:Lamb.18}
\begin{array}{l}
\displaystyle
\frac{\omega^2}{\omega^2_0}\,(c'_1\,\cos\omega t\,+\,9\,c'_3\,\cos 3\omega t)\,\simeq\,\\
\\
\simeq\,
2\,J_1(\vartheta_0)\,\cos\omega t\,-\,2\,J_3(\vartheta_0)\,\cos 3\omega t\,,
\end{array}
\end{equation}
where also the frequency $\omega$ has to be re-calculated by using again  Eq.~(\ref{Eq:Lamb.14.1}) written as
\begin{equation}
\label{Eq:Lamb.19}
\displaystyle
c'_1\,+\,c'_3\,\simeq\,\vartheta_0\,,
\end{equation}
so that, after some algebra, we obtain
\begin{equation}
\label{Eq:Lamb.20}
\left\{
\begin{array}{l}
\displaystyle
c'_1\,\simeq\,\frac {\vartheta_0}{1\,+\,\eta}\,,\\
\\
\displaystyle
c'_3\,\simeq\,\frac {\vartheta_0}{1\,+\,\displaystyle\frac 1\eta}\,=\,\eta\,c'_1\,,\\
\\
\displaystyle
\frac{\omega^2}{\omega^2_0}\,\simeq\,\frac{2\,{J_1(\vartheta_0)}}{\vartheta_0}\,(1\,+\,\eta)\,,
\end{array}
\right.
\end{equation}
where the correction term $\eta$ is given by
\begin{equation}
\label{Eq:Lamb.21}
\eta\,=\,-\,\displaystyle\frac 19\,\displaystyle\frac{J_3(\vartheta_0)}{J_1(\vartheta_0)}\,.
\end{equation}

\begin{figure}[!ht]
 \begin{minipage}[b]{6.cm}
 \centerline{
   \includegraphics[width=6cm]{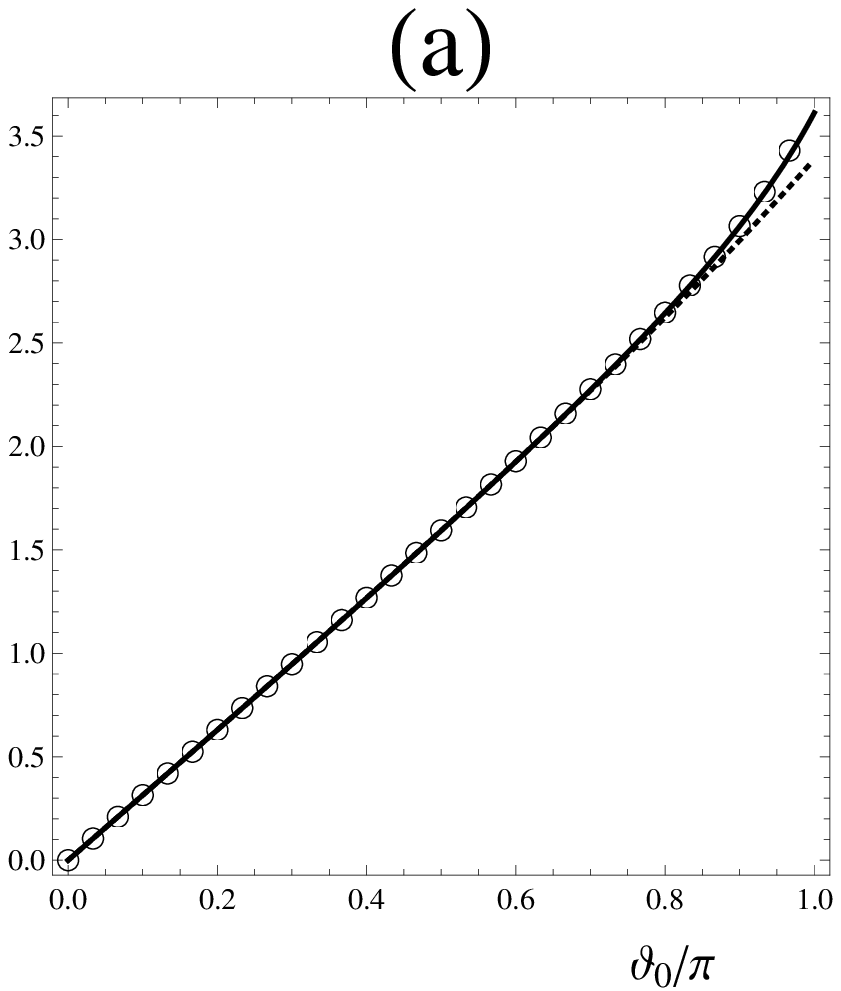}%{Fig.C1.First.eps}
   }
 \end{minipage}
 \hspace*{5mm}
 \begin{minipage}[b]{6.cm}
   \centering
   \includegraphics[width=6cm]{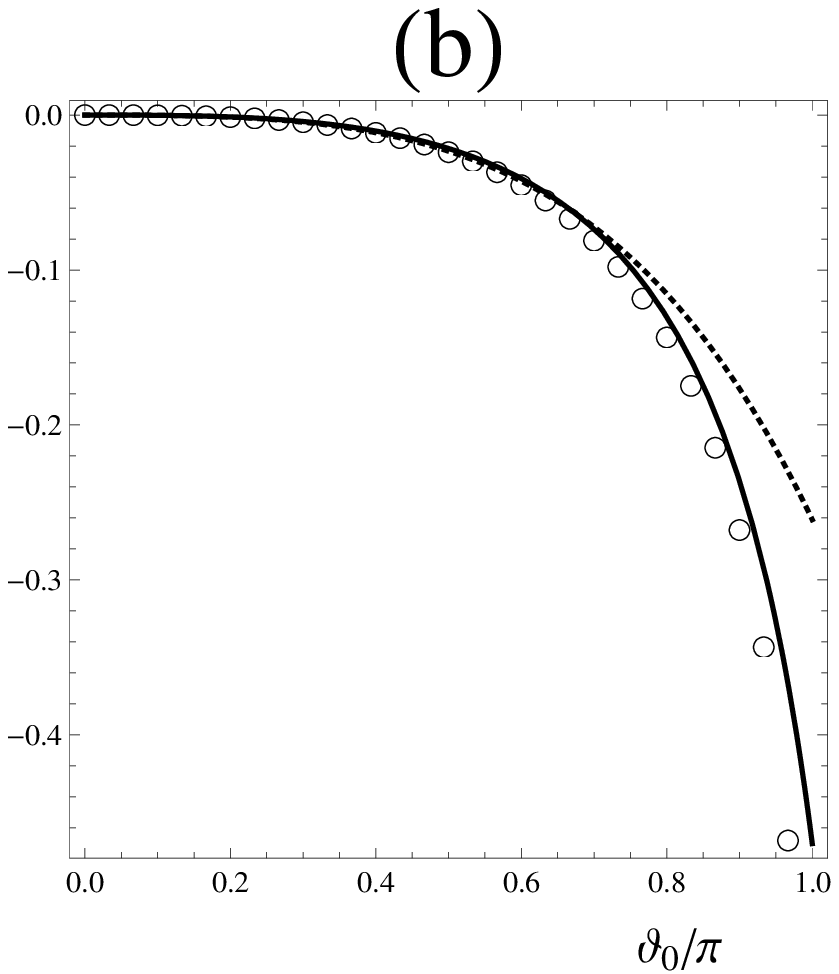}%{Fig.C3.First.eps}
 \end{minipage}
\caption{The same as in Fig.~\ref{Fig:Lamb.0}, but for the first-order solution.
(a): Behaviour of the coefficient $c'_1$. (b): Behaviour of the coefficient $c'_3$.}
\label{Fig:Lamb.2.1}
\end{figure}

Similarly as done in Fig.~\ref{Fig:Lamb.0}, Fig.~\ref{Fig:Lamb.2.1}
shows the behaviour of the expanding coefficients given in Eq.~(\ref{Eq:Lamb.20})
as functions of the initial amplitude $\theta_0$.
Moreover, Fig.~\ref{Fig:Lamb.2.1} shows the same comparison reported
in Fig.~\ref{Fig:Lamb.2}, which now displays an agreement with the exact solution that is really good for the cases depicted in figures~(a), (b), and (c), but it is still far from being satisfactory for the case of figure~(d). 
\begin{figure}[!ht]
 \begin{minipage}[b]{6cm}
   \centering
   \includegraphics[width=6cm]{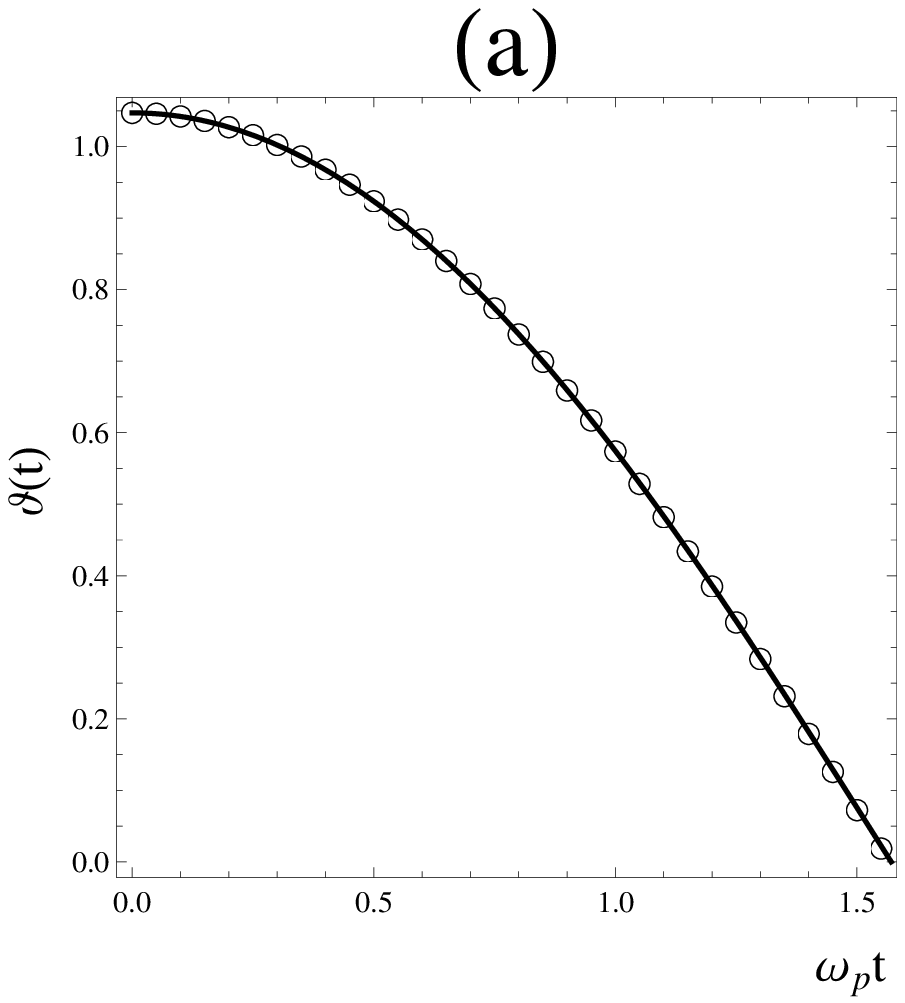} %{Fig.Comparison.FirstA.eps}
   \includegraphics[width=6cm]{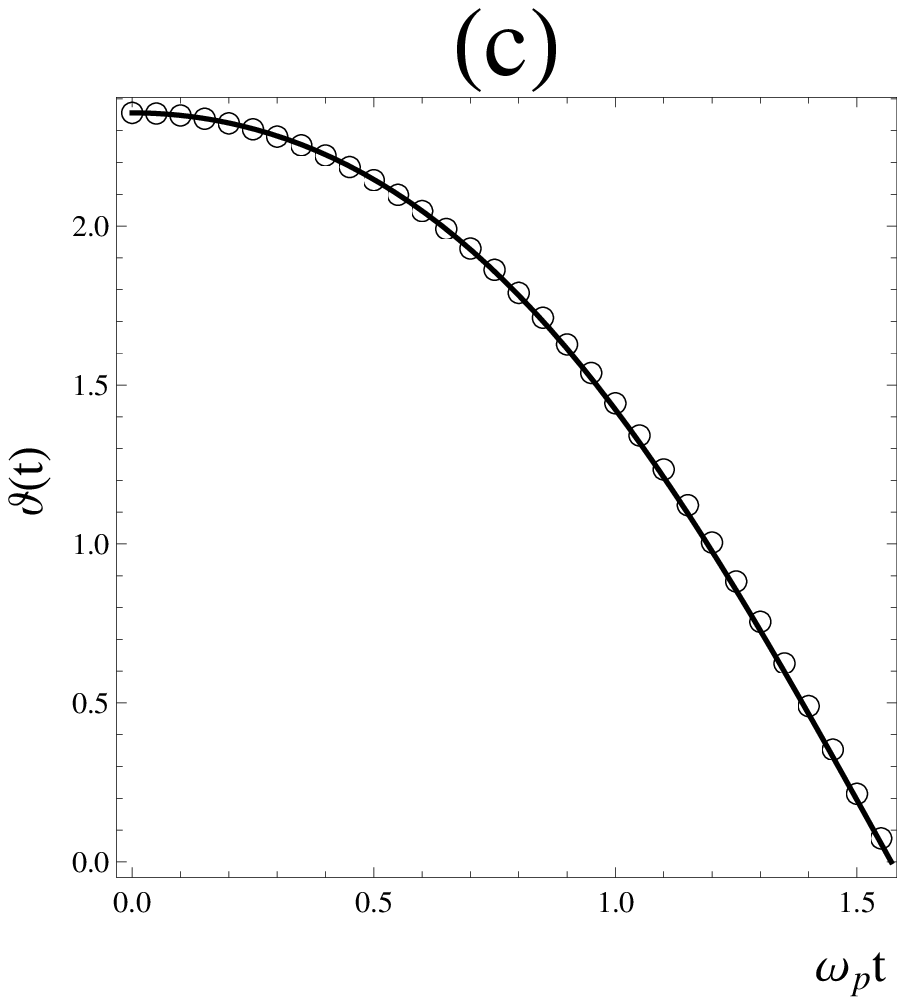} %{Fig.Comparison.FirstC.eps}
 \end{minipage}
 \ \hspace{2mm} \hspace{3mm} \
 \begin{minipage}[b]{6cm}
  \centering
   \includegraphics[width=6cm]{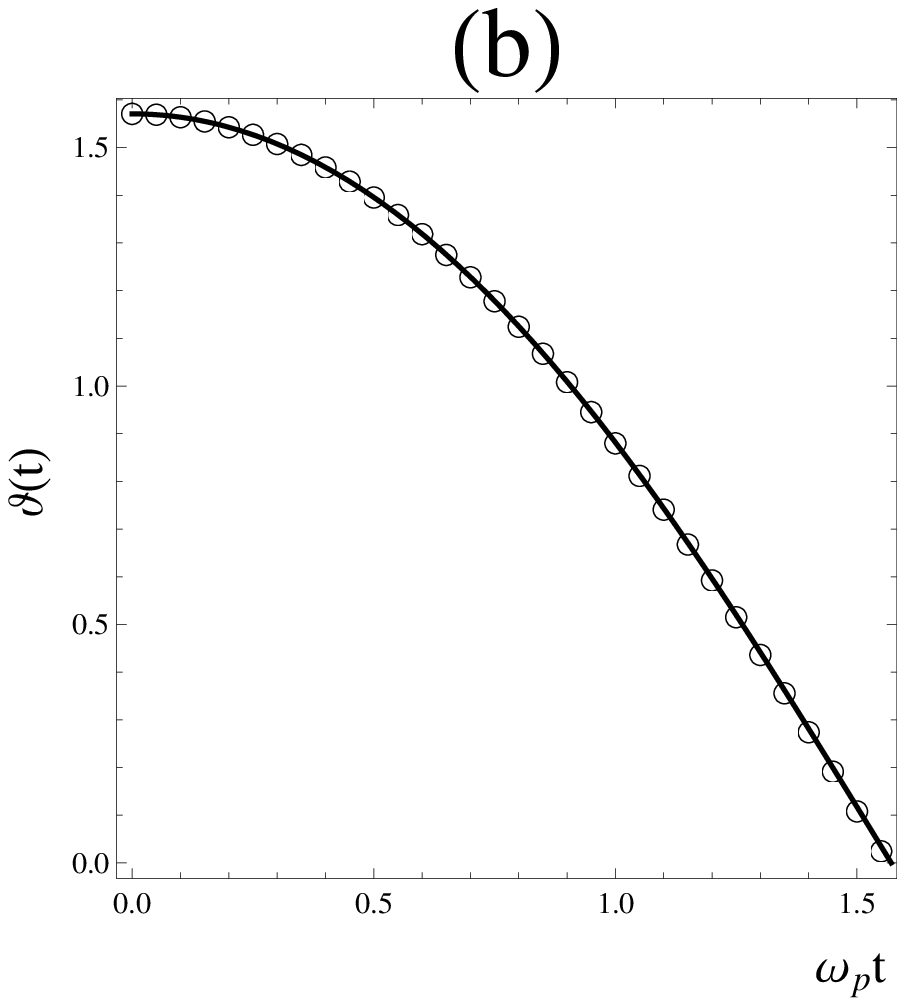} %{Fig.Comparison.FirstB.eps}
   \includegraphics[width=6cm]{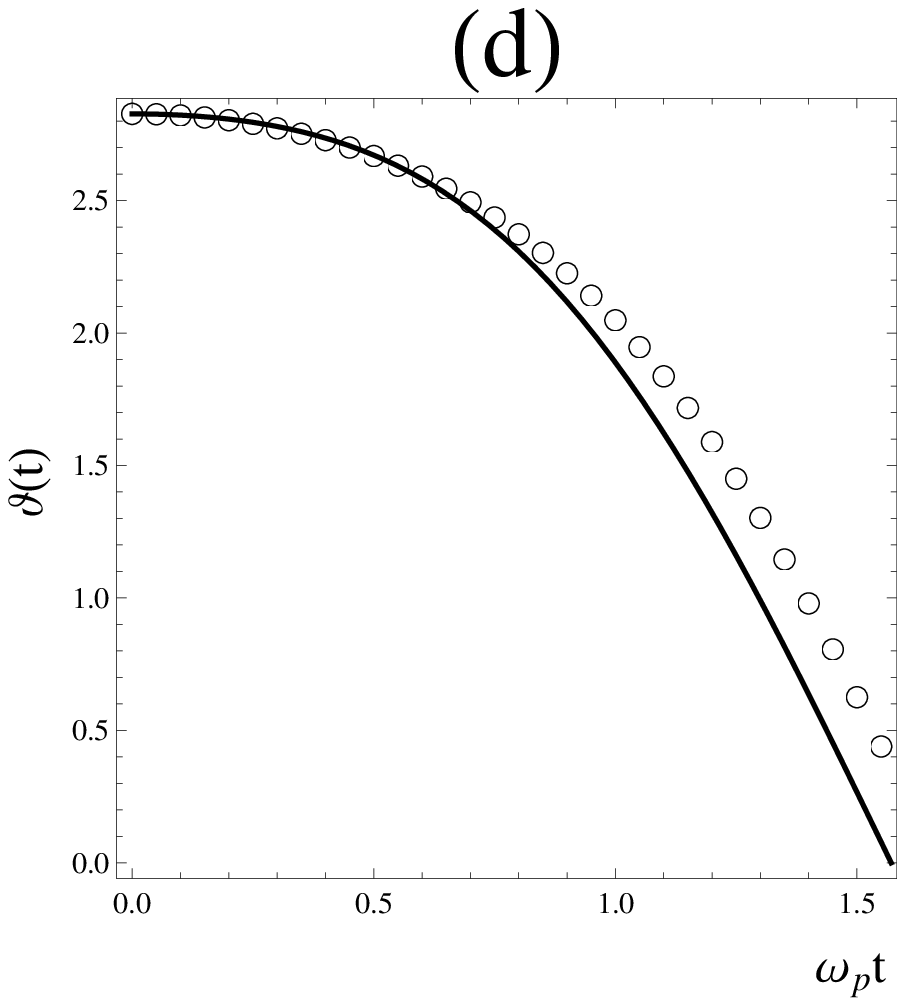} %{Fig.Comparison.FirstD.eps}
 \end{minipage}
\caption{The same as in Fig.~\ref{Fig:Lamb.2} but with the first-order solution in place of the lowest-order one.}
\label{Fig:Lamb.3}
\end{figure}
The reason is due to the fact that for the latter case the pendulum starts from an initial position close to the vertical, which leads to a law of motion with a ``flat'' shape in the space $(t,\vartheta)$ (this is more evident for values 
$\vartheta_0$ which are closer to $\pi$) and, accordingly, it would require a greater number of harmonics to achieve a better representation. We prefer not to detail here the whole development of the second-order solution, which is confined in the~\ref{App:SecondOrderSolution} for interested readers, but to limit
ourselves to note that it will be obtained by keeping in the Fourier series~(\ref{Eq:Lamb.12}) the first three terms whose expanding coefficient estimates,
say $\{c''_1,c''_3,c''_5\}$ are shown, as functions of $\vartheta_0$, in Fig.~\ref{Fig:Lamb.3.1}(a), (b), and (c), respectively.
\begin{figure}[!ht]
 \begin{minipage}[b]{4.3cm}
 \centerline{
   \includegraphics[width=4.3cm]{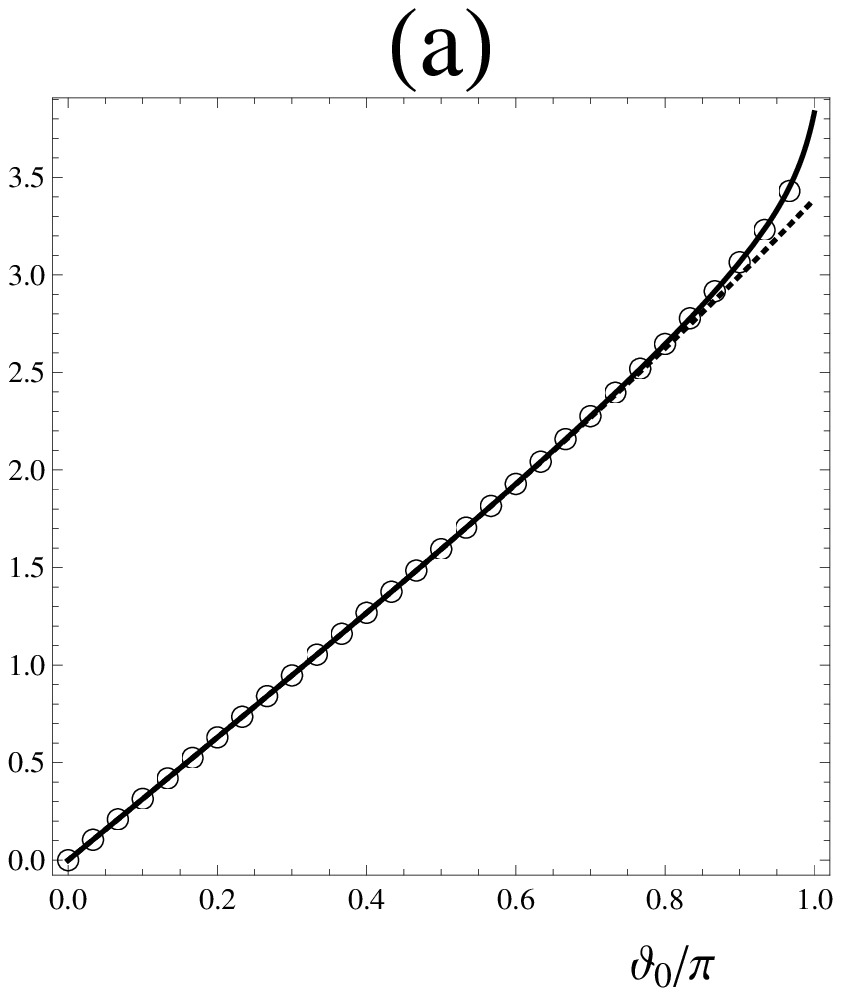}%{Fig.C1.Second.eps}
   }
 \end{minipage}
% \hspace*{5mm}
 \begin{minipage}[b]{4.3cm}
 \centerline{
   \includegraphics[width=4.3cm]{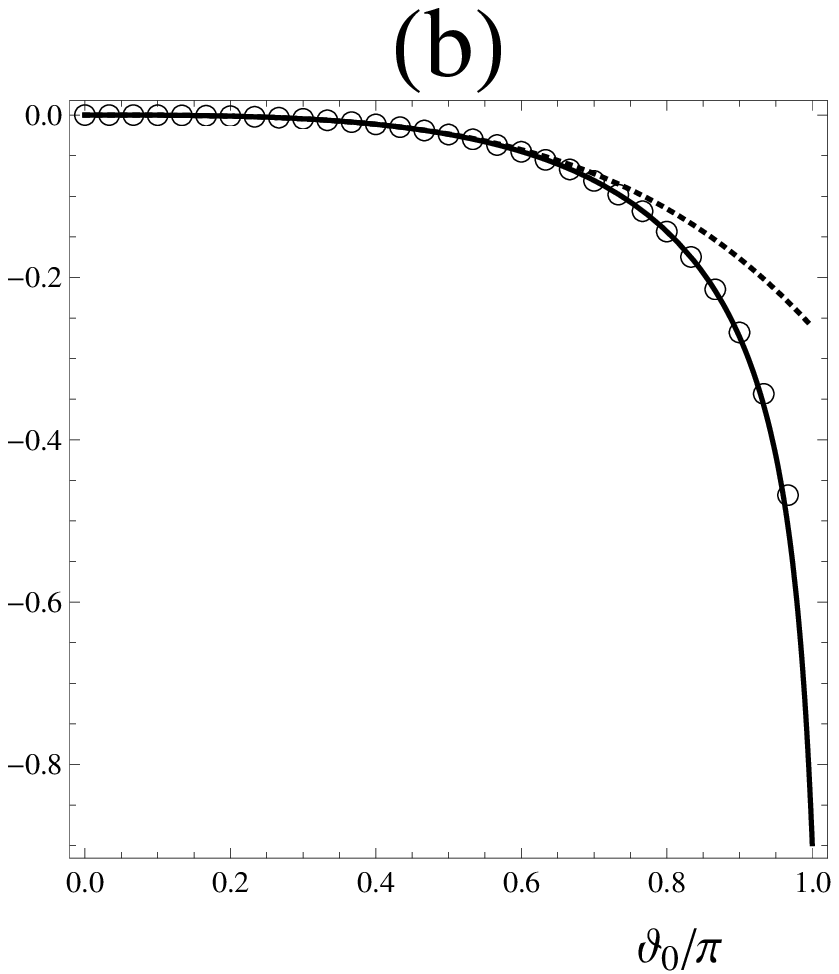}%{Fig.C3.Second.eps}
   }
 \end{minipage}
% \hspace*{5mm}
 \begin{minipage}[b]{4.3cm}
 \centerline{
   \includegraphics[width=4.3cm]{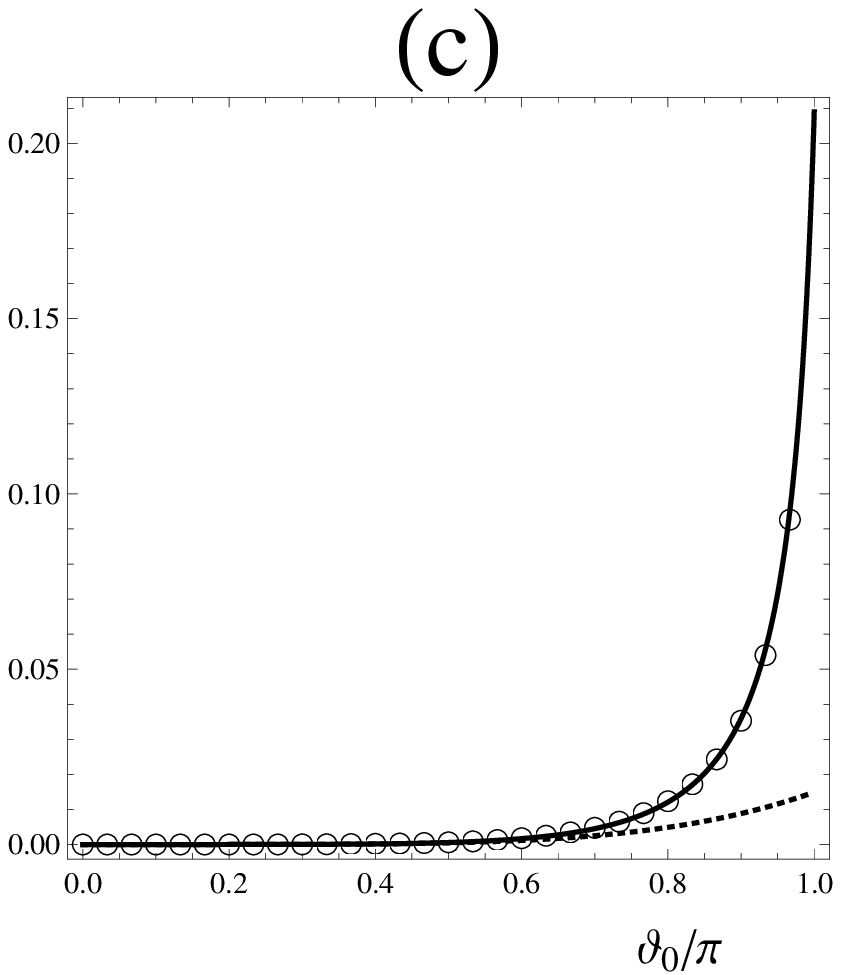}%{Fig.C5.Second.eps}
   }
 \end{minipage}
\caption{The same as in Fig.~\ref{Fig:Lamb.0}, but for the second-order solution.
(a): Behaviour of the coefficient $c''_1$. (b): Behaviour of the coefficient $c''_3$. (c): Behaviour of the coefficient $c''_5$.}
\label{Fig:Lamb.3.1}
\end{figure}
Figure~\ref{Fig:Lamb.4} shows the results obtained by using the second-order 
solution, which includes the first three harmonics in the Fourier expansion
of Eq.~(\ref{Eq:Lamb.12}). In particular figure~(a) corresponds to 
the choice $\vartheta_0=9\pi/10$, in order for a visual comparison with 
Fig.~3(d) to be allowed. It is seen that the second-order solution displays an excellent agreement with the exact solution considerably better than that 
corresponding to the first-order one. However, for initial positions closer to the vertical also the second-order solution reveals to be inadequate, as shown in figures~(b) and~(c), so that one should pass to the third-order solution, and so on. 

To conclude the present section we wish to compare the estimates of the
pendulum period, obtained at the above described three subsequent approximation levels, with the exact expression. This is shown in Fig.~\ref{Fig:Lamb.5}(a), where the exact behavior of $T/T_0$ as a function of $\theta_0$ is shown (open circles) together with the estimate provided by the lowest-order (dashed curve), the first-order (dotted curve), and the second-order (solid curve) approximants 
of the law of motion. To give a more quantitative information, in Fig.~\ref{Fig:Lamb.5}(b)
the corresponding values of the relative error (in percentage) are also plotted. 
\begin{figure}[!ht]
\begin{minipage}[b]{4.2cm}
	\centering
	\includegraphics[width=4.2cm]{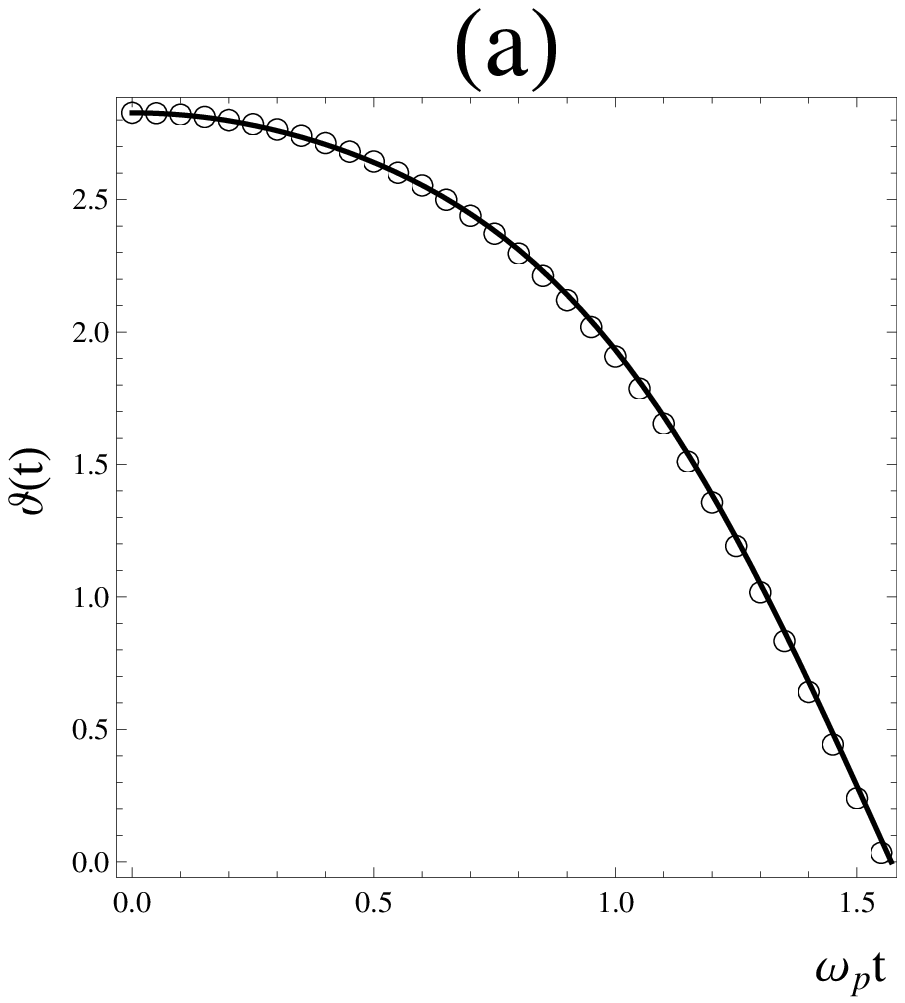}%{Fig.Lamb.4a.eps}
 \end{minipage}
\begin{minipage}[b]{4.2cm}
	\centering
	\includegraphics[width=4.2cm]{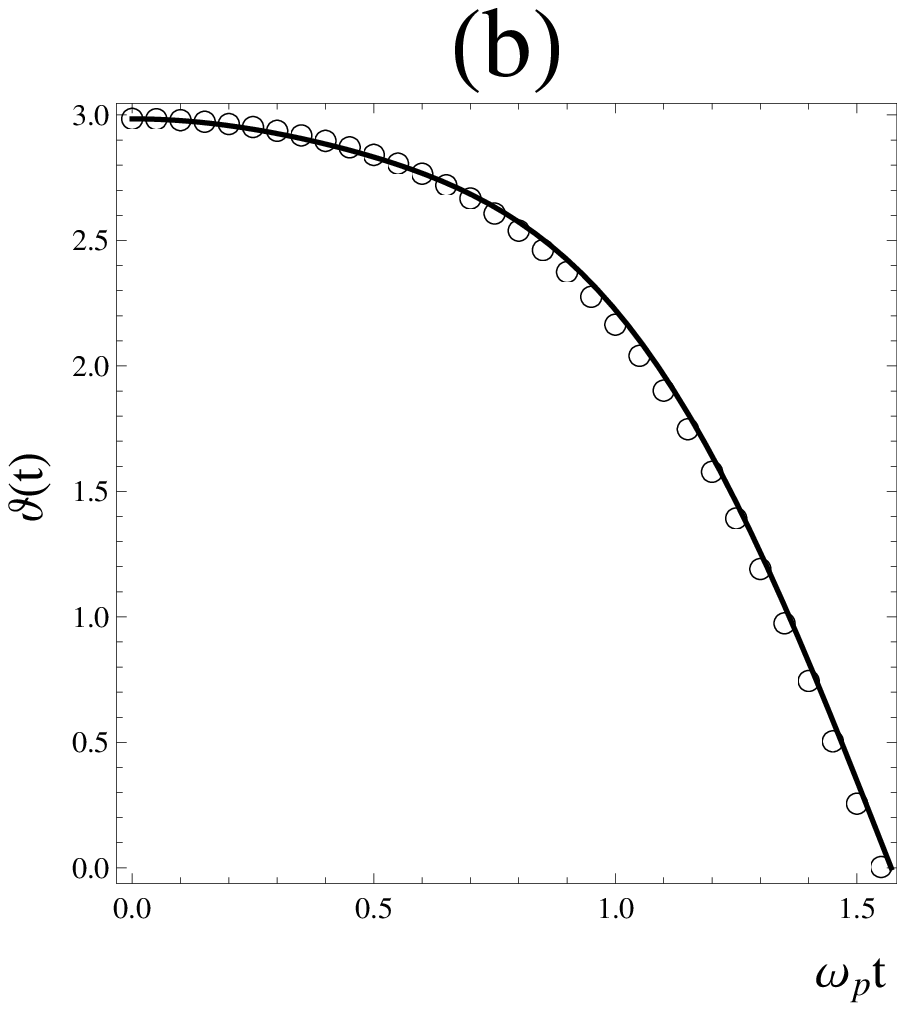}%{Fig.Lamb.4b.eps}
 \end{minipage}
\begin{minipage}[b]{4.2cm}
	\centering
	\includegraphics[width=4.2cm]{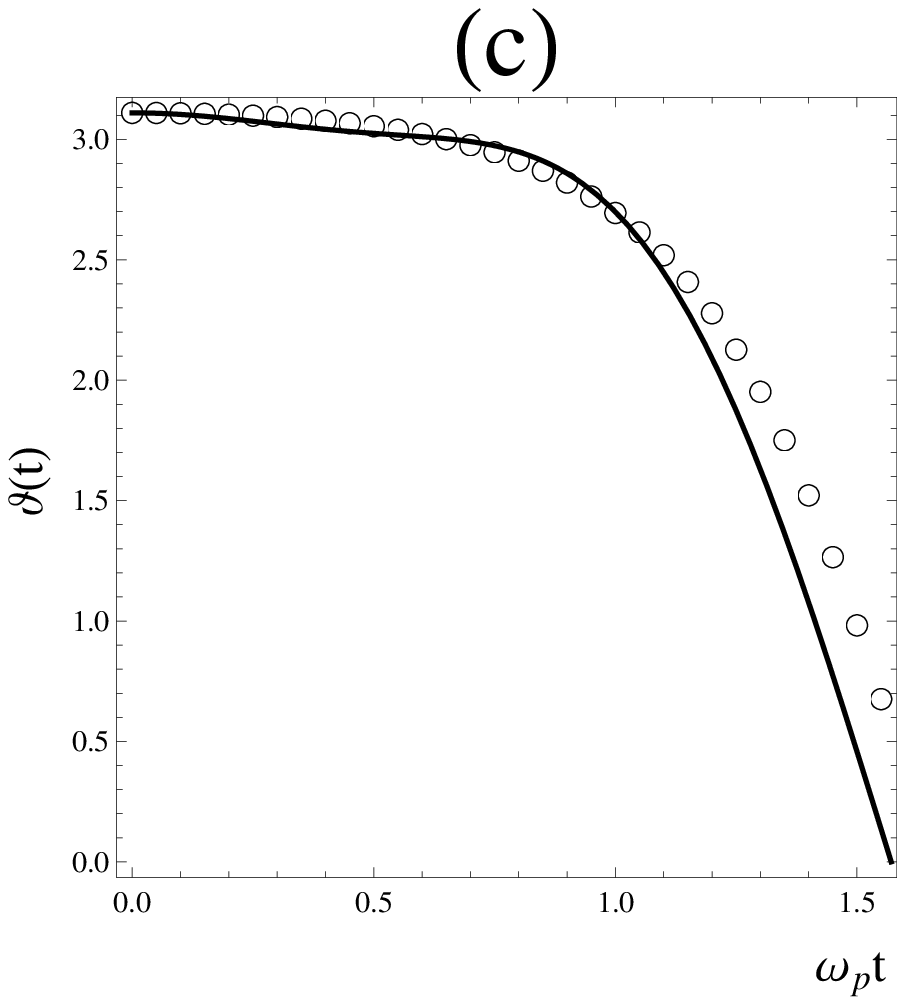}%{Fig.Lamb.4c.eps}
 \end{minipage}
\caption{
Comparison between the temporal behaviours of the exact 
(open circles) law of motion and  the second-order estimate~(solid curve)
for $\vartheta_0=9\pi/10$ (a), $95\pi/100$ (b), and $99\pi/100$ (c).}
\label{Fig:Lamb.4}
\end{figure}
\begin{figure}[!ht]
 \begin{minipage}[b]{6.cm}
 \centerline{
   \includegraphics[width=6cm]{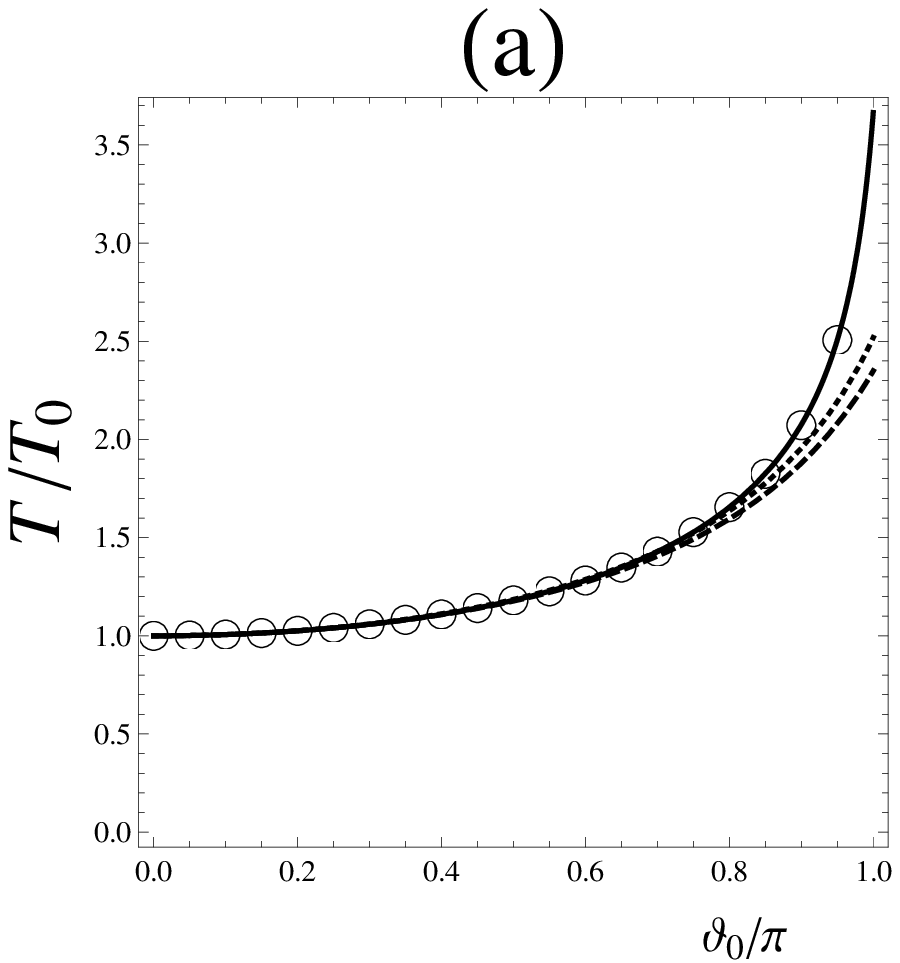}%{Fig.Comparison.PeriodA.eps}
   }
 \end{minipage}
 \hspace*{5mm}
 \begin{minipage}[b]{6.cm}
   \centering
   \includegraphics[width=6cm]{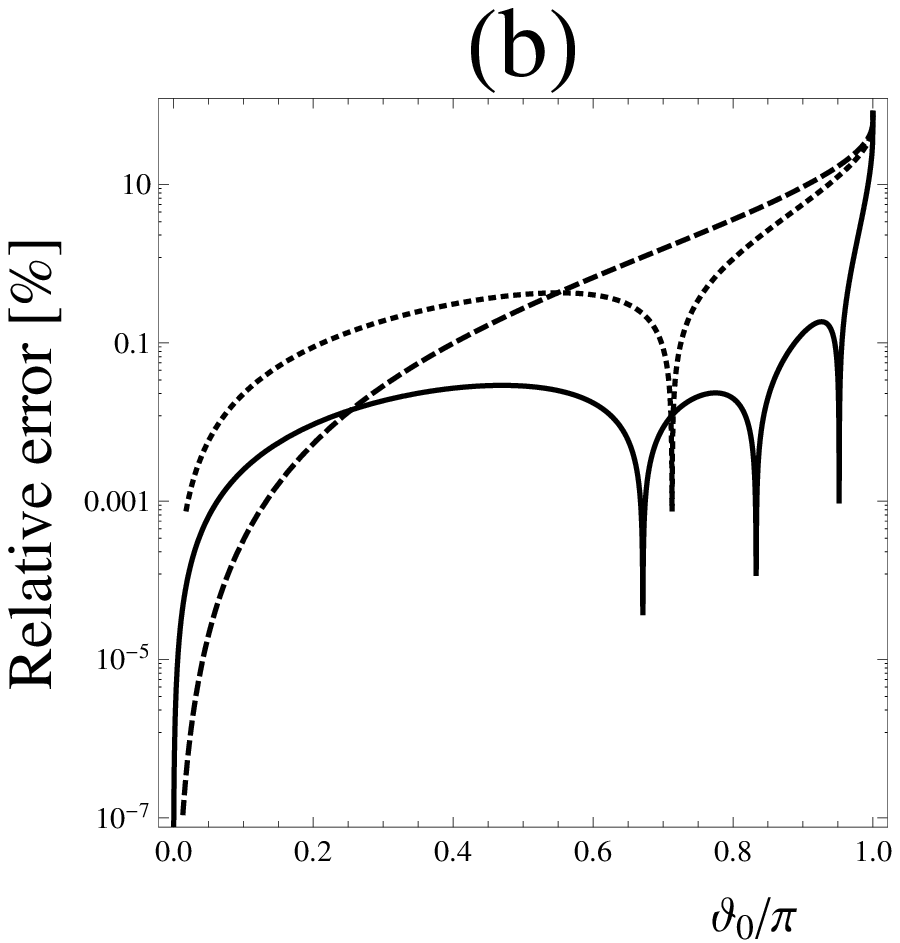}%{Fig.Comparison.PeriodB.eps}
 \end{minipage}
\caption{(a): exact behavior of $T/T_0$ as a function of $\theta_0$ 
(open circles), together with that corresponding to the lowest-order 
(dashed curve), the first-order (dotted curve), and the second-order 
(solid curve) solutions. (b): corresponding values of the relative error 
(in percentage). 
 }
\label{Fig:Lamb.5}
\end{figure}

\section{Conclusions}
\label{Sec:Conclusions}

Nonlinear physics is a fascinating world where it is not always possible to establish general rules similarly to what happens for most linear problems, that can often be solved through ``closed-form approaches.'' The study of the simple pendulum dynamics is not an exception and it is really difficult to find elementary ways to approach it and to offer its solution to undergraduates. 
An approach to the solution of the simple pendulum problem which should appear 
to be ``natural'' is to exploit the periodicity of the temporal law of motion by representing it as a Fourier series, with unkown expanding coefficients that 
should be found via the constraints imposed by the nonlinear differential equation and the appropriate initial conditions. In the present paper such an approach has been used to build up analytical approximants of the pendulum solution, as well as of the period, in terms of suitable superpositions of Bessel functions of the first kind. In particular, at the lowest approximation level the pendulum period turns out to be well described (with accuracies better than 0.2 \%for oscillation amplitudes up to $\pi/2$) by the so-called ``besinc'' function, well known in physical optics. 
We also believe that, from a mere didactical point of view, such a connection could be exploited to convince even an audience of skeptic students about the fact that different phenomena in physics are often, depending on the approximation level,  quantitatively described by equations having similar mathematical aspects~\cite{Feynman/Lectures/II}. Accordingly, methods aimed at finding their solutions must necessarily share a common signature. This, in particular, is true as far as the key role played by some classes of so-called ``special functions'' (and the Bessel family is one of the most important)
is concerned, a fact that makes them of fundamental importance in applications to several different branches of physics~\cite{Berry/2001}.

\section*{Acknowedgments}

I wish to thank Turi Maria Spinozzi for his help during the preparation
of the manuscript.

\appendix

\section{Derivation of the second-order solution}
\label{App:SecondOrderSolution}

Again we have to start from Eq.~(\ref{Eq:Lamb.14}) written in the following form:
\begin{equation}
\label{Eq:SecondOrderSolution.1}
\begin{array}{l}
\displaystyle
\frac{\omega^2}{\omega^2_0}
\,(c''_1\,\cos\omega t\,+\,9\,c''_3\,\cos 3\omega t\,+\,25\,c''_5\,\cos 5\omega t))\,\simeq\,\\
\\
\simeq\,
\sin(c'_1\,\cos\omega t\,+\,c'_3\,\cos 3\omega t)\,,
\end{array}
\end{equation}
where $\{c'_1,c'_3\}$  denotes the set of coefficients found at the first level and given in Eqs.~(\ref{Eq:Lamb.20}) and~(\ref{Eq:Lamb.21}), while the triplet~$\{c''_1,\,c''_3,\,c''_5\}$ denotes the set of coefficients that we have to include in the representation of $\vartheta(t)$ at the second level of expansion. To find them consider the Fourier series expansion of the function $\sin\vartheta(t)$ \emph{at the first level}, namely
\begin{equation}
\label{Eq:SecondOrderSolution.3}
\begin{array}{l}
\displaystyle
\sin(c'_1\,\cos \omega t\,+\,c'_2\,\cos 3\omega t)\,=\,\\
\\
=\,
%\sum_{k=0}^\infty\,
C_{1}\,\cos\,\cos \omega t\,+\,
C_{3}\,\cos\,\cos 3\omega t\,+\,
C_{5}\,\cos\,\cos 5\omega t\,+\,
\ldots\,,
\end{array}
\end{equation}
which will be truncated up to the first three terms 
before inserting into Eq.~(\ref{Eq:SecondOrderSolution.1})
with the expanding coefficients $\{C_{1},C_3,C_5\}$ being given by
\begin{equation}
\label{Eq:SecondOrderSolution.4}
\begin{array}{l}
\displaystyle
C_{2k+1}\,=\,
\frac 1\pi\,
\int_0^{2\pi}\,
\cos\,[(2k+1)x]\,
\sin(c'_1\,\cos x\,+\,c'_3\,\cos 3x)\,.
\end{array}
\end{equation}
Although the above integrals cannot be expressed exactly in closed form, analytical estimates of them can be obtained by approximating the integrand
as follows:
\begin{equation}
\label{Eq:SecondOrderSolution.4.1}
\begin{array}{l}
\displaystyle
\sin(c'_1\,\cos x\,+\,c'_3\,\cos 3x)\,\simeq\,
\sin(c'_1\,\cos x)\,+\,c'_3\,\cos 3x\,\cos(c'_1\,\cos x)\,,
\end{array}
\end{equation}
which, once substituted into Eq.~(\ref{Eq:SecondOrderSolution.4}), after 
some algebra gives
\begin{equation}
\label{Eq:SecondOrderSolution.4.1.1}
\begin{array}{lcl}
C_{2k+1} & \simeq &
\displaystyle
\frac 1\pi\,
\int_0^{2\pi}\,
\cos\,[(2k+1)x]\,
\sin(c'_1\,\cos x)\,\\
&&\\
& + & 
\displaystyle
\frac {c'_3}\pi\,
\int_0^{2\pi}\,
\cos\,[(2k+1)x]\,\cos\,3x\,
\cos(c'_1\,\cos x)\,,
\end{array}
\end{equation}
and, on taking formulas~(10.12.2) and~(10.12.3) of~\cite{DLMF}
into account,
\begin{equation}
\label{Eq:SecondOrderSolution.4.2}
\begin{array}{lcl}
(-1)^k\,C_{2k+1} & \simeq &
\displaystyle
2\,J_{2k+1}(c'_1)
%&&\\
%& + & 
\displaystyle
\,+\,{c'_3}\,
[J_{2k+4}(c'_1)\,-\,J_{2k-2}(c'_1)]\,.
\end{array}
\end{equation}
Finally, on substituting from Eq.~(\ref{Eq:SecondOrderSolution.3}) into
Eq.~(\ref{Eq:SecondOrderSolution.1}) and on comparing both sides of the resulting equation term by term, the following system is obtained:
\begin{equation}
\label{Eq:SecondOrderSolution.5}
\begin{array}{l}
\displaystyle
\frac{\omega^2}{\omega^2_0}\,c''_{2k+1}\,=\,
\frac {C_{2k+1}}{(2k+1)^2}\,,\qquad\,k\,=\,0,1,2\,,
\end{array}
\end{equation}
which, together with the boundary condition
\begin{equation}
\label{Eq:SecondOrderSolution.6}
\begin{array}{l}
\displaystyle
c''_1\,+\,c''_3\,+\,c''_5\,=\,\vartheta_0\,,
\end{array}
\end{equation}
and on taking Eq.~(\ref{Eq:SecondOrderSolution.4.2}) into account, allows both the frequency $\omega$ and the triplet~$\{c''_1,\,c''_3,\,c''_5\}$
to be straightforwardly found. 

Of course the above described procedure can be further iterated to retrieve,
in principle, trigonometric approximants of the pendulum law of motion 
up to arbitrarily high accuracies, that is left to the interested reader.

%%%%%%%%%%%%%%%%%%%%%%%%%%%%%%%%%%%%%%%%%%%%%%%%%%%%%%%%%%%%%%%

\end{document}